\documentstyle[epsfig,12pt,a4p]{article}

\newcommand{\pipipi}{\mbox{$\pi^+\pi^-\pi^0$ }}

\begin{document}
\begin{titlepage}
\def\footnoterule{\hrule width 1.0\columnwidth}
\begin{tabbing}
put this on the right hand corner using tabbing so it looks
 and neat and in \= \kill
\> {5 January 1998}
\end{tabbing}
\bigskip
\bigskip
\begin{center}{\Large  {\bf A study of the
centrally produced \pipipi channel
in pp interactions at 450 GeV/c}
}\end{center}
\bigskip
\bigskip
\begin{center}{        The WA102 Collaboration
}\end{center}\bigskip
\begin{center}{
D.\thinspace Barberis$^{  5}$,
W.\thinspace Beusch$^{   5}$,
F.G.\thinspace Binon$^{   7}$,
A.M.\thinspace Blick$^{   6}$,
F.E.\thinspace Close$^{  4}$,
K.M.\thinspace Danielsen$^{ 12}$,
A.V.\thinspace Dolgopolov$^{  6}$,
S.V.\thinspace Donskov$^{  6}$,
B.C.\thinspace Earl$^{  4}$,
D.\thinspace Evans$^{  4}$,
B.R.\thinspace French$^{  5}$,
T.\thinspace Hino$^{ 13}$,
S.\thinspace Inaba$^{   9}$,
A.V.\thinspace Inyakin$^{  6}$,
T.\thinspace Ishida$^{   9}$,
A.\thinspace Jacholkowski$^{   5}$,
T.\thinspace Jacobsen$^{  12}$,
G.V.\thinspace Khaustov$^{  6}$,
T.\thinspace Kinashi$^{  11}$,
J.B.\thinspace Kinson$^{   4}$,
A.\thinspace Kirk$^{   4}$,
W.\thinspace Klempt$^{  5}$,
V.\thinspace Kolosov$^{  6}$,
A.A.\thinspace Kondashov$^{  6}$,
A.A.\thinspace Lednev$^{  6}$,
V.\thinspace Lenti$^{  5}$,
S.\thinspace Maljukov$^{   8}$,
P.\thinspace Martinengo$^{   5}$,
I.\thinspace Minashvili$^{   8}$,
K.\thinspace Myklebost$^{   3}$,
T.\thinspace Nakagawa$^{  13}$,
K.L.\thinspace Norman$^{   4}$,
J.M.\thinspace Olsen$^{   3}$,
J.P.\thinspace Peigneux$^{  1}$,
S.A.\thinspace Polovnikov$^{  6}$,
V.A.\thinspace Polyakov$^{  6}$,
Yu.D.\thinspace Prokoshkin$^{\dag  6}$,
V.\thinspace Romanovsky$^{   8}$,
H.\thinspace Rotscheidt$^{   5}$,
V.\thinspace Rumyantsev$^{   8}$,
N.\thinspace Russakovich$^{   8}$,
V.D.\thinspace Samoylenko$^{  6}$,
A.\thinspace Semenov$^{   8}$,
M.\thinspace Sen\'{e}$^{   5}$,
R.\thinspace Sen\'{e}$^{   5}$,
P.M.\thinspace Shagin$^{  6}$,
H.\thinspace Shimizu$^{ 14}$,
A.V.\thinspace Singovsky$^{  6}$,
A.\thinspace Sobol$^{   6}$,
A.\thinspace Solovjev$^{   8}$,
M.\thinspace Stassinaki$^{   2}$,
J.P.\thinspace Stroot$^{  7}$,
V.P.\thinspace Sugonyaev$^{  6}$,
K.\thinspace Takamatsu$^{ 10}$,
G.\thinspace Tchlatchidze$^{   8}$,
T.\thinspace Tsuru$^{   9}$,
G.\thinspace Vassiliadis$^{\dag   2}$,
M.\thinspace Venables$^{  4}$,
O.\thinspace Villalobos Baillie$^{   4}$,
M.F.\thinspace Votruba$^{   4}$,
Y.\thinspace Yasu$^{   9}$.
}\end{center}

\begin{center}{\bf {{\bf Abstract}}}\end{center}

{
The reaction
$ pp \rightarrow p_{f} (\pi^{+}\pi^{-}\pi^{0}) p_{s}$
has been studied at 450 GeV/c
in an experiment designed to search for gluonic states.
A spin analysis has been performed and the $dP_T$ filter applied. The
analysis confirms the previous observation that all undisputed
$q \overline q$ states are suppressed at small $dP_T$.
In addition, a clear difference is observed in the production
mechanism for the $\eta$ and $\omega$.
}
\bigskip
\bigskip
\bigskip
\bigskip\begin{center}{{Submitted to Physics Letters}}
\end{center}
\bigskip
\bigskip
\begin{tabbing}
aba \=   \kill
$^\dag$ \> \small
Deceased. \\
$^1$ \> \small
LAPP-IN2P3, Annecy, France. \\
$^2$ \> \small
Athens University, Nuclear Physics Department, Athens, Greece. \\
$^3$ \> \small
Bergen University, Bergen, Norway. \\
$^4$ \> \small
School of Physics and Astronomy, University of Birmingham, Birmingham, U.K. \\
$^5$ \> \small
CERN - European Organization for Nuclear Research, Geneva, Switzerland. \\
$^6$ \> \small
IHEP, Protvino, Russia. \\
$^7$ \> \small
IISN, Belgium. \\
$^8$ \> \small
JINR, Dubna, Russia. \\
$^9$ \> \small
High Energy Accelerator Research Organization (KEK), Tsukuba, Ibaraki 305,
Japan. \\
$^{10}$ \> \small
Faculty of Engineering, Miyazaki University, Miyazaki, Japan. \\
$^{11}$ \> \small
RCNP, Osaka University, Osaka, Japan. \\
$^{12}$ \> \small
Oslo University, Oslo, Norway. \\
$^{13}$ \> \small
Faculty of Science, Tohoku University, Aoba-ku, Sendai 980, Japan. \\
$^{14}$ \> \small
Faculty of Science, Yamagata University, Yamagata 990, Japan. \\
\end{tabbing}
\end{titlepage}
\setcounter{page}{2}
\bigskip
\par
The WA102 collaboration
has recently described the application of a kinematical filter
on centrally produced events~\cite{WADPT}.
This filter, which
has been proposed as a glueball-$q \overline q$ filter~\cite{ck97},
was originally based on the idea of
Double Pomeron Exchange (DPE) and hence it is interesting to
observe the effect of this filter on channels that cannot
be produced by DPE.
This paper presents new results from the WA102 experiment on
the centrally produced
$\pi^{+}\pi^{-}\pi^{0}$ final state in the reaction
\begin{equation}
pp \rightarrow p_{f} (\pi^{+}\pi^{-}\pi^{0}) p_{s}
\label{eq:a}
\end{equation}
at 450 GeV/c.
The subscripts f and s indicate the
fastest and slowest particles in the laboratory respectively.
Due to the G parity of the \pipipi system it cannot be produced by
DPE.
These data represent an increase
by a factor of fifty in the previously published
WA76 results on the
centrally produced \pipipi final state~\cite{WA763pi}.
\par
The data come from experiment WA102
which has been performed using the CERN Omega Spectrometer.
The layout of the Omega Spectrometer used in this run is similar to that
described in ref.~\cite{wa9192} with the replacement of the
OLGA calorimeter by GAMS~4000~\cite{gams}.
\par
Reaction~(\ref{eq:a})
has been isolated from the sample of events having four outgoing
charged tracks plus two $\gamma$s 
reconstructed in the electromagnetic
calorimeter\footnote{The showers associated with the impact of
the charged tracks on the calorimeter
have been removed from the event before the
requirement of only two $\gamma$s was made.}
by first imposing the following cuts on the components of the
missing momentum:
$|$missing~$P_{x}| <  17.0$ GeV/c,
$|$missing~$P_{y}| <  0.16$ GeV/c and
$|$missing~$P_{z}| <  0.12$ GeV/c,
where the x axis is along the beam
direction.
A correlation between
pulse-height and momentum
obtained from a system of
scintillation counters was used to ensure that the slow
particle was a proton.
\par
The effective mass of the two $\gamma$s shows
a clear $\pi^0$ signal ($\sigma$~=~17~MeV),
which was selected by requiring
0.1~$<$~m($\gamma \gamma$)~$<$~0.17~GeV.
The quantity $\Delta$, defined as
$ \Delta = MM^{2}(p_{f}p_{s}) - M^{2}(\pi^{+}\pi^{-}\pi^{0})$,
where $MM^{2}(p_{f}p_{s})$ is the missing mass squared of the two
outgoing protons,
was then calculated for each event and
a cut of $|\Delta|$ $\leq$ 3.0 (GeV)$^{2}$ was used to select the
$\pi^{+}\pi^{-}\pi^{0}$
channel. Events containing a fast $\Delta^{++}(1232) $
were removed if $M(p_{f} \pi^{+}) < 1.3 $ GeV, which left
1~323~136 centrally produced \pipipi events.
After the selection of the \pipipi channel a kinematical fit was performed in
order to apply overall energy and momentum balance.
\par
Fig.~\ref{fi:a}a) shows the acceptance corrected
$\pi^{+}\pi^{-}\pi^{0}$
effective mass spectrum renormalised to the total number of observed events.
In addition to clear $\eta(547)$ and $\omega(782)$
signals there is a broad enhancement
at 1.2~GeV, some evidence for the $a_2(1320)$ and another broad enhancement
at 1.67~GeV.
\par
A Dalitz plot analysis
of the \pipipi final state has been performed
using Zemach tensors and a standard isobar model~\cite{ABRAM}.
The analysis has assumed $\rho(770)$, $\sigma$, $f_0(980)$ and $f_2(1270)$
intermediate states
and that
only relative angular momenta up to 2 contribute.
For the $\rho \pi$ decays both I~=~0 and I~=~1 final states have been
considered.
The $\sigma$ stands for the $\pi\pi$ S-wave amplitude squared,
and the parameterisation of
Zou and
Bugg~\cite{re:zbugg} has been used in this analysis.
\par
The geometrical acceptance of the apparatus has also been evaluated over
the Dalitz plot of the \pipipi system
in 40~MeV intervals between 0.8 and 2.0 GeV.
In order to perform a spin parity analysis the
log likelihood function, ${\cal L}_{j}=\sum_{i}\log P_{j}(i)$,
is defined by combining the probabilities of all events in 40 MeV
\pipipi mass bins from 0.84 to 2.0 GeV.
In order to include more than one wave in the fit
the incoherent sum of various
event fractions $a_{j}$ is calculated:
\begin{equation}
{\cal L}=\sum_{i}\log \left(\sum_{j}a_{j}P_{j}(i) +
(1-\sum_{j}a_{j})\right)
\end{equation}
where the term
$(1-\sum_{j}a_{j})$ represents the phase space background
which is a free parameter in each bin.
The negative log likelihood function ($-{\cal L} $) is then minimised using
MINUIT~\cite{re:MINUIT}. Coherence between different $J^{P}$ states
and between different isobar amplitudes of a given $J^{P}$ have
been neglected in the fit.
Different combinations of waves and isobars have been tried and
insignificant contributions have been removed from the final fit.
Using Monte Carlo simulations it has been found that the feed-through
from one spin parity to another is negligible and that the
peaks in the spin analysis cannot be produced by phase space
or acceptance effects.
The fit generates the phase space background as that part of the data
not associated with a given wave
on a bin by bin basis and one requirement is that this
background is a smoothly varying function that shows no residual
resonance structure.
\par
The result of the best fit is shown in
fig.~\ref{fi:b} for the total data sample.
A fit using
$J^{PC}=1^{++}$~$\rho^\pm\pi^\mp $ S-wave,
$J^{PC}=2^{++}$~$\rho^\pm\pi^\mp $ D-wave,
$J^{PC}=2^{-+}$~$\rho^\pm\pi^\mp $ P-wave,
$J^{PC}=2^{-+}$~$f_2(1270) \pi^0$ S-wave,
$J^{PC}=1^{--}$~$\rho \pi$ P-wave
and phase space is found to be sufficient in order to describe the data.
The phase space distribution resulting from the fit shows no remaining
resonance structure.
\par
The $J^{PC}=1^{++}$~$\rho^\pm\pi^\mp $ S-wave
shows a broad enhancement. It has been fitted using
an expression of the form:
\begin{equation}
\frac{dN}{dm} = \frac{m_{a_1} \Gamma_{a_1}(m)}
{(m^2 - m^2_{a_1})^2 + \Gamma^2_{a_1}(m) m^2_{a_1}}
\end{equation}
where $\Gamma_{a_1}(m)$ is a mass dependent width of the form
\begin{equation}
\Gamma_{a_1}(m) = \Gamma_{a_1} m_{a_1} \frac{\rho^{1^+S}(m)}
{\rho^{1^+S}(m_{a_1})}
\end{equation}
where $\rho^{1^+S}(m)$  is the phase space for S-wave
$\rho^\pm(770) \pi^\mp$,
including the effects of the mass dependent $\rho^\pm (770)$ width and
of interference between the two $\rho^\pm (770)$  bands in the Dalitz
plot~\cite{BOWLER}.
The results of the fit are given in table~\ref{ta:a} and the
mass and width are
consistent with the PDG values for the $a_1(1260)$~\cite{PDG96}.
\par
The $J^{PC}=2^{++}$~$\rho^\pm\pi^\mp $ D-wave
shows a peak in the $a_2(1320)$ region and has been
fitted using a relativistic spin 2 Breit-Wigner.
The results of the fit are given in table~\ref{ta:a}.
The
$J^{PC}=2^{-+}$~$\rho^\pm\pi^\mp $ P-wave and
$J^{PC}=2^{-+}$~$f_2(1270) \pi^0$ S-wave have similar distributions
consistent with the $\pi_2(1670)$.
The $J^{PC}=2^{-+}$~$\rho^\pm\pi^\mp $ P-wave has been fitted with a
spin 1 relativistic Breit-Wigner and the
$J^{PC}=2^{-+}$~$f_2(1270) \pi^0$ S-wave has been fitted with a
spin 0 relativistic Breit-Wigner. The masses and widths are similar
and are given in table~\ref{ta:a}.
The
$J^{PC}=1^{--}$~$\rho^\pm\pi^\mp $ P-wave is dominated by a peak
at the mass of the $\phi(1020)$ which has been fitted with a
spin 1 relativistic Breit-Wigner convoluted with a Gaussian
to describe the experimental resolution.
\par
The \pipipi mass spectrum shown in
fig.~\ref{fi:a}a) has been fitted with the resonances described above
together with a Gaussian~($\sigma$~=~11MeV) to describe
the $\eta$, a Breit-Wigner
convoluted with a Gaussian~($\sigma$~=~18~MeV) to describe the $\omega$ and a
background of the form
$a(m-m_{th})^b$~exp$(-cm-dm^2-em^3)$, where
$m$ is the \pipipi mass,
$m_{th}$ is the threshold mass and
a,b,c,d,e are fit parameters. The fit is found to describe the data
well and yields masses for the $\eta$ and $\omega$ given in
table~\ref{ta:a}.
\par
The branching ratio for the $\pi_2(1670)$ to $\rho \pi$ and
$f_2(1270) \pi$ has been calculated taking into account the unseen decay modes
of the $f_2(1270)$
and gives
\begin{equation}
\frac{\pi_2(1670) \rightarrow \rho \pi}{\pi_2(1670) \rightarrow f_2(1270) \pi}
= 0.57  \pm 0.02 ^{+ 0.03}_{-0.02}
\end{equation}
which is in good agreement
with the PDG values of 0.55~$\pm$~0.08~\cite{PDG96}.
The systematic error quoted
has three major sources. The first comes from
the error in the branching ratio of $f_2(1270)$ to
$\pi \pi$~\cite{PDG96}.
The second is due to
the
determination of the number of events in each decay mode;
this has been estimated by comparing the ratio calculated from the total
data sample compared to the ratio calculated
from the fits performed as a function of $dP_T$ (see below).
Finally estimates of possible backgrounds below the signals have been made
for both decay modes.
\par
In previous analyses of other channels it has been observed that
when the centrally produced system has been analysed
as a function of the parameter $dP_T$, which is the difference
in the transverse momentum vectors of the two exchange particles~\cite{WADPT},
all the undisputed $q \overline q$ states are suppressed at small
$dP_T$.
Therefore a study of
the \pipipi system has been made as a function of $dP_T$.
The \pipipi mass spectrum is presented in
fig.~\ref{fi:a}b), c) and d)
for $dP_T$~$\leq$~0.2~GeV,
0.2~$\leq$~$dP_T$~$\leq$~0.5~GeV and
$dP_T$~$\geq$~0.5~GeV respectively.
In order to determine the dependence of the $\eta$ and the $\omega$
as a function of $dP_T$, a fit has been made to these mass spectra using
the parameters from the fit to the total mass spectrum.
For the $a_1(1260)$, $a_2(1320)$, $\pi_2(1670)$ and $\phi(1020)$ the
spin analysis has been performed as a function of $dP_T$ and the results
are shown in fig.~\ref{fi:c}.
The $J^{PC}=2^{-+}$~$\rho^\pm\pi^\mp $ P-wave and
$J^{PC}=2^{-+}$~$f_2(1270) \pi^0$ S-wave have been summed together to produce
a single $J^{PC}=2^{-+}$ wave.
The waves have been fitted with the Breit-Wigners described above and
the results are given in
table~\ref{ta:b}.
As can be seen
all the resonances are suppressed at small $dP_T$. However, it is
interesting to note that they do not all have the same
$dP_T$ behaviour.
\par
The azimuthal angle ($\phi$) between the fast and slow proton
is related to $dP_T$ by
\begin{center}
$cos \phi  = \frac{ P_T^2 - dP_T^2 }{4 P_{T_1} P_{T_2}}$
\end{center}
where $P_T$ is the transverse momentum of the central system and
$P_{T_{1,2}}$ is the transverse momentum of the exchanged particle.
{}From table~\ref{ta:b} it can be seen that
the $\eta$ and
the $\omega$ have a similar $dP_T$ dependence; however,
they have a dramatically
different $\phi$ dependence. Fig.~\ref{fi:4}a) and b) show the
angle $\phi$ for 0.5~$\le$~m(\pipipi)~$\le$~0.6 GeV ($\eta$ region) and
for 0.735~$\le$~m(\pipipi)~$\le$~0.835 GeV ($\omega$ region) respectively.
This suggests that the $\eta$ and
the $\omega$  have very different production mechanisms.
Further evidence for this comes from studying the
four momentum transfer squared from one of the proton vertices
which is shown in
fig.~\ref{fi:4}c) and d) for the $\eta$ and $\omega$ respectively.
The $|t|$ distribution for the $\omega$ is more steep than that for the
$\eta$ and extends to $|t|$~=~0 unlike the $\eta$ distribution which
appears to turn over.
\par
In order to investigate the centre of mass energy dependence
of $\eta$ and $\omega$ production, the ratio of the acceptance corrected number
of $\eta$s and $\omega$s has been calculated
using data from the WA76 experiment which was performed
at an incident beam energy of 85~GeV/c
($\sqrt s$~=~12.7~GeV)~\cite{WA763pi} and
from this current experiment ($\sqrt s $~=~29~GeV).
\begin{equation}
\frac{\sigma(\eta)}{\sigma(\omega)}\thinspace  =\thinspace
0.20\pm0.02\thinspace (\sqrt s~=~12.7~\mathrm{GeV})\thinspace  =\thinspace
0.09~\pm~0.01\thinspace (\sqrt s ~=~29~\mathrm{GeV})
\end{equation}
which again indicates a different production mechanism for the two resonances.
\par
In summary, a study of the centrally produced \pipipi system shows
that the most prominent resonance signals are due to the $\eta$, $\omega$,
$\phi(1020)$, $a_1(1260)$, $a_2(1320)$ and $\pi_2(1670)$.
All these states are found to be suppressed at small $dP_T$.
In addition,
a clear difference is observed in the production characteristics of the
$\eta$ compared to the $\omega$.

\newpage
{ \large \bf Tables \rm}
\begin{table}[h]
\caption{Parameters of resonances in the fit to the
$\pi^{+}\pi^{-}\pi^{0}$ mass spectrum and waves.}
\label{ta:a}
\vspace{1in}
\begin{center}
\begin{tabular}{|c|c|c|c|c|c|} \hline
 & & & & & \\
 &Mass (MeV) & Width (MeV) &Observed & $I(J^{PC})$ &Number of\\
 & & & decay mode & & events\\
 & & & & & \\ \hline
 & & & & & \\
$\eta$  & 549 $\pm$ 1 &  -  & &$0(0^{-+})$ &21590$\pm$644 \\
 & & & & & \\ \hline
 & & & & & \\
$\omega(782)$ & 785 $\pm$ 2 & 8.4 (fixed)  &  & $0(1^{--})$ &248511$\pm$2118\\
 & & & & & \\ \hline
 & & & & & \\
$\phi(1020)$ &1021 $\pm$ 4 & 4 (fixed) & $\rho\pi$ & $0(1^{--})$
&9053$\pm$490\\
 & & & & & \\ \hline
 & & & & & \\
$a_{1}(1260)$ &1240 $\pm$ 10 & 400 $\pm$ 35 & $\rho\pi$ & $1(1^{++})$
&327957$\pm$2678\\
 & & & & & \\ \hline
 & & & && \\
$a_{2}(1320)$ &1317 $\pm$ 3 &120 $\pm$ 10 &$\rho\pi$ &$1(2^{++})$
&35616$\pm$648\\
 & & & & & \\ \hline
 & & & && \\
$\pi_{2}(1670)$ &1669 $\pm$ 4 &268 $\pm$ 15 &$\rho\pi$ &$1(2^{-+})$
&23535$\pm$598\\
                &1670 $\pm$ 4 &256 $\pm$ 15 &$f_2(1270)\pi$ &$1(2^{-+})$
&23692$\pm$549\\
 & & & & & \\ \hline
\end{tabular}
\end{center}
\end{table}
\newpage
\begin{table}[h]
\caption{Resonance production as a function of $dP_T$
expressed as a percentage of its total contribution.}
\label{ta:b}
\vspace{1in}
\begin{center}
\begin{tabular}{|c|c|c|c|} \hline
 & & &  \\
 &$dP_T$$\leq$0.2 GeV & 0.2$\leq$$dP_T$$\leq$0.5 GeV &$dP_T$$\geq$0.5 GeV\\
 & & & \\ \hline
 & & & \\
$\eta$  &5.3 $\pm$ 0.5 $\pm$ 0.5 & 42 $\pm$ 1 $\pm 1 $ &52 $\pm$ 1 $\pm$ 1 \\
 & & & \\ \hline
 & & & \\
$\omega(782)$  &13 $\pm$ 1 $\pm$ 1 & 44 $\pm$ 1 $\pm$ 1 &43 $\pm$ 1 $\pm$ 1 \\
 & & & \\ \hline
 & & & \\
$\phi(1020) $  &8 $\pm$ 2 $\pm$ 2 & 47 $\pm$ 3 $\pm$ 3 &45 $\pm$ 2 $\pm$ 4\\
 & & & \\ \hline
 & & & \\
$a_{1}(1260)$  &17 $\pm$ 1 $\pm$ 3 & 53 $\pm$ 3 $\pm$ 2&29 $\pm$ 2 $\pm$ 2 \\
 & & & \\ \hline
 & & & \\
$a_{2}(1320)$  &4 $\pm$ 2 $\pm$ 3 & 37 $\pm$ 2 $\pm$ 1 &59 $\pm$ 3 $\pm$ 3 \\
 & & & \\ \hline
 & & & \\
$\pi_{2}(1670)$  &15 $\pm$ 1 $\pm$ 1 & 51 $\pm$ 2 $\pm$ 3 &33 $\pm$ 1 $\pm$ 3
\\
 & & & \\ \hline
\end{tabular}
\end{center}
\end{table}
\newpage
{ \large \bf Figures \rm}
\begin{figure}[h]
\caption{The
$\pi^{+}\pi^{-}\pi^{0}$
effective mass spectrum a) for the total data with fit using
3 Breit-Wigners
b) for $dP_T$~$\leq$~0.2~GeV,
c) for 0.2~$\leq$~$dP_T$~$\leq$~0.5~GeV and
d) for $dP_T$~$\geq$~0.5~GeV.}
\label{fi:a}
\end{figure}
\begin{figure}[h]
\caption{Results of the spin parity analysis.
The superimposed curves are the resonance contributions coming from
the fits described in the text.}
\label{fi:b}
\end{figure}
\begin{figure}[h]
\caption{Results of the spin parity analysis as a function of $dP_T$.
}
\label{fi:c}
\end{figure}
\begin{figure}[h]
\caption{The azimuthal angle ($\phi$) between the two outgoing protons
for the a) $\eta$ and b) $\omega$.
The four momentum transfer squared ($|t|$) from one of the proton
vertices
for the c) $\eta$ and d) $\omega$.
}
\label{fi:4}
\end{figure}
\newpage
\begin{center}
\epsfig{figure=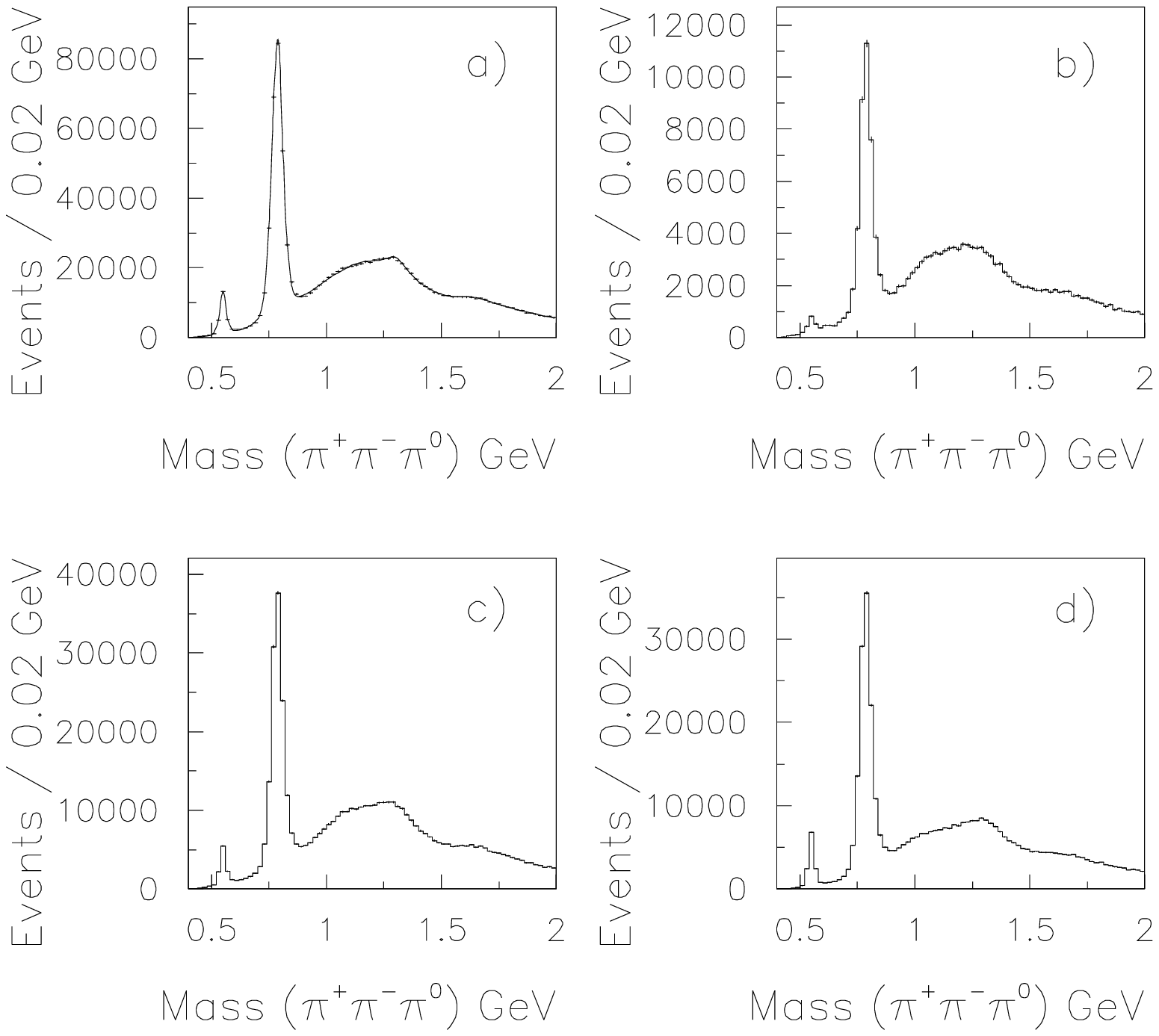,height=22cm,width=17cm}
\end{center}
\begin{center} {Figure 1} \end{center}
\newpage
\begin{center}
\epsfig{figure=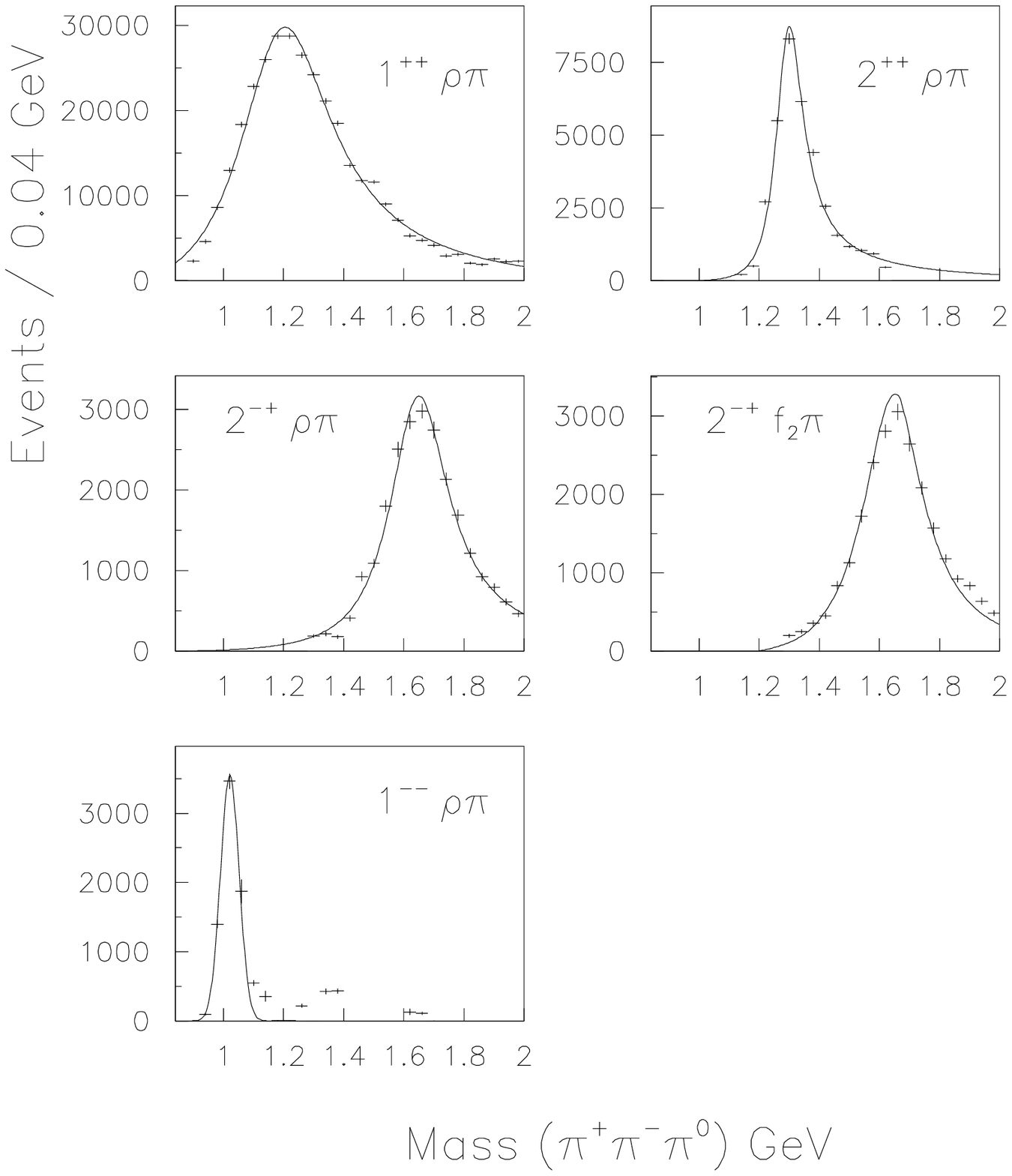,height=22cm,width=17cm}
\end{center}
\begin{center} {Figure 2} \end{center}
\newpage
\begin{center}
\epsfig{figure=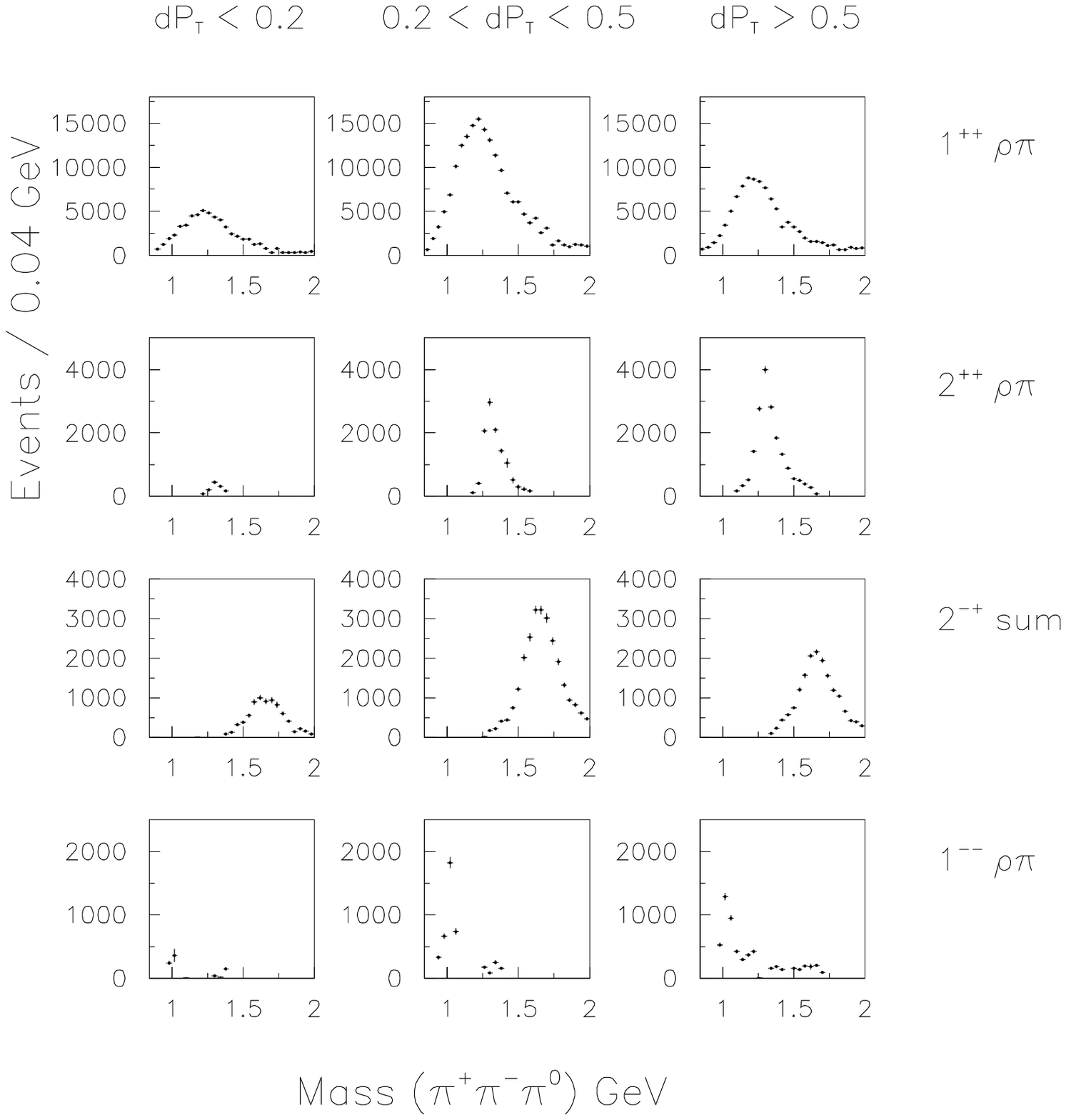,height=22cm,width=17cm}
\end{center}
\begin{center} {Figure 3} \end{center}
\newpage
\begin{center}
\epsfig{figure=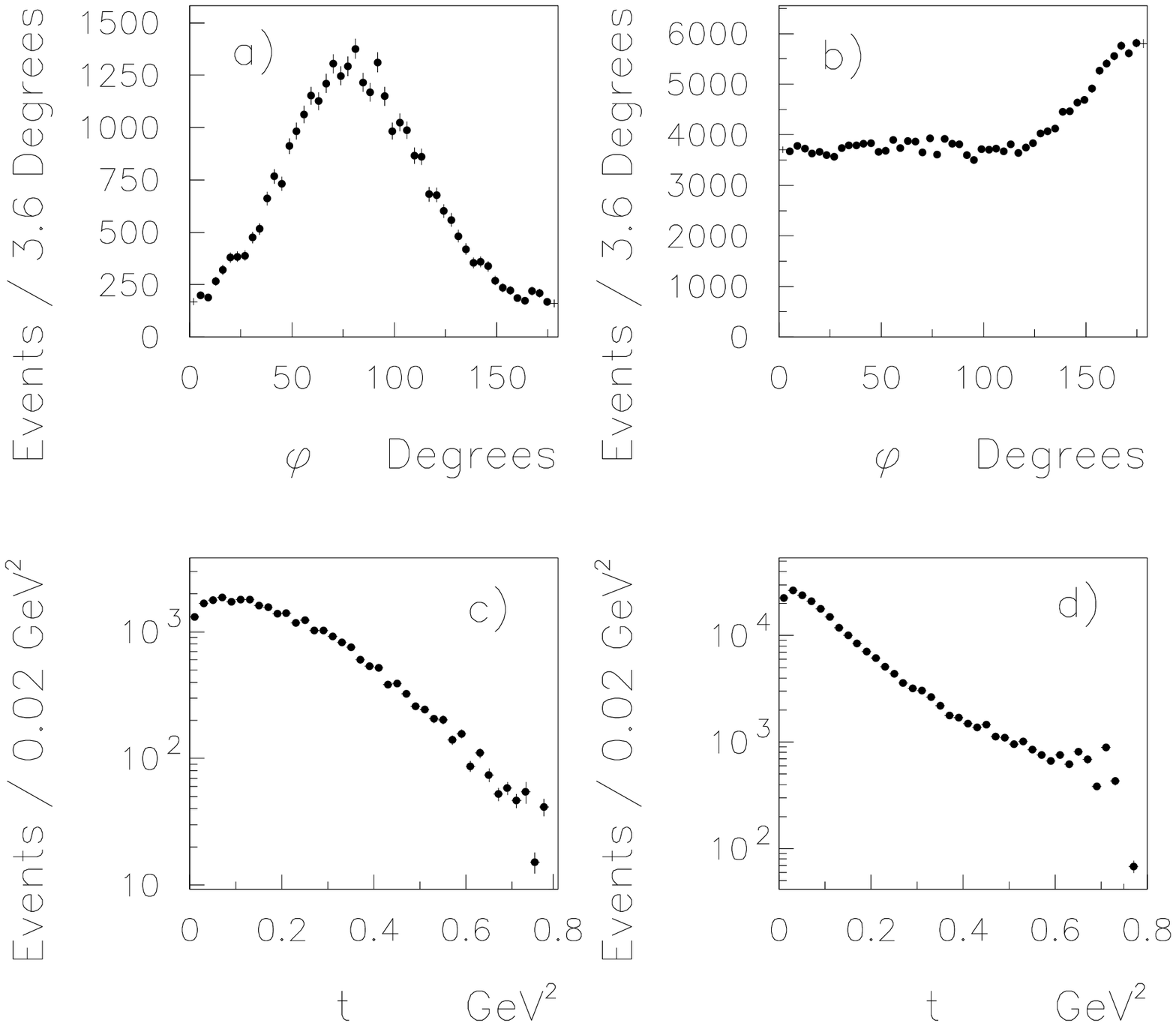,height=22cm,width=17cm}
\end{center}
\begin{center} {Figure 4} \end{center}

\bigskip
\newpage
\begin{thebibliography}{99}
\bibitem{WADPT}
D. Barberis {\em et al.,} Phys. Lett. {\bf B397 } \rm (1997) 339.
\bibitem{ck97}
F.E. Close and A. Kirk, Phys. Lett. {\bf B397 } \rm (1997) 333.
\bibitem{WA763pi}
T.A. Armstrong {\em et al.,} Zeit. Phys. {\bf C48 } \rm (1990) 213.
\bibitem{wa9192}
F. Antinori {\em et al.,} Il Nuovo Cimento {\bf A107 } (1994) 1857.
\bibitem{gams}
D. Alde {\em et al.,} Nucl. Phys. {\bf B269} (1986) 485.
\bibitem{ABRAM}
M. Abramovich {\em et al.,} Nucl. Phys. {\bf B23} (1970) 466.
\bibitem{re:zbugg}
B. S. Zou and D. V. Bugg, Phys. ReV. {\bf D48} (1993) R3948.
\bibitem{re:MINUIT}
F. James and M. Roos, MINUIT Computer Physics Communications
{\bf 10 } \rm (1975) 343; CERN-D506 (1989).
\bibitem{BOWLER}
M.G. Bowler, Phys. Lett. {\bf B182} (1986) 400.
\bibitem{PDG96}
Particle Data Group, Phys Rev. {\bf D54} (1996) 1.
\end{thebibliography}
\end{document}